\documentstyle[aps,multicol,prb,epsf]{revtex} 
 
\begin{document} 
\draft

\title{Weak localization correction to the FS interface resistance} 
 
\author{Edward McCann$\sp{1}$, Vladimir I. Fal'ko$\sp{1}$, A. F. Volkov$\sp{1,2}$,
and C. J. Lambert$\sp{1}$} 
\address{$\sp{1}$ School of Physics and Chemistry, Lancaster University,
Lancaster, LA1 4YB, UK}
\address{$\sp{2}$ Institute for Radio-Engineering and Electronics,
Moscow, Russia
} 
\date{\today} 
\maketitle \begin{abstract} {
The classical resistance of a contact between a ferromagnet (F) and a
superconductor
(S) acquires an additional contact term as compared to the contact between
a ferromagnet and a normal metal.
The necessity to match spin-polarized current in a ferromagnet to spin-less
current in the superconductor results in the accumulation of
non-equilibrium polarization near the F/S interface.
In the present work, we show that the weak localization correction to the
classical diffusion coefficient, $\delta D$, is dependent on the degree
of polarization, with majority spins more likely to be reflected from the
interface than minority spins.
Taking into account the change in the spin polarized particle distribution
in the F wire arising from $\delta D$, we calculate the weak localization
correction to the F/S contact resistance.
}\end{abstract} \pacs{PACS numbers: 
73.23.-b, 
72.10.Bg, 
74.80.Dm, 
}

\begin{multicols}{2}

\bibliographystyle{simpl1}
%
%
\section{Introduction}
%

Novel phenomena in mesoscopic systems have been the subject of intense
theoretical and experimental interest for
many years\cite{soh97,imr97}.
Observed effects such as weak localization and universal conductance
fluctuations are due to the quantum interference of electrons
at low temperatures.
An important branch of mesoscopic physics is that of hybrid
nanostructures where the influence of a superconductor (S) on a
phase coherent normal (N) region is studied\cite{Bee97,L+R98}.
At low temperatures, the role of Andreev reflection\cite{And64} is
important, whereby particles in the N region with excitation energies
$\varepsilon$
lower than the superconducting gap energy $\Delta$ are reflected
from the N/S interface as holes.
Studies of subgap transport have shown that superconducting condensate
penetration into the N region, known as the proximity effect,
may substantially change the resistance of
an N/S circuit.

Lately, improvements in the microprocessing of metals has led to
the fabrication of nanostructures combining ferromagnetic (F) and
S materials\cite{Pet94,L+G96,Gir98,Vas97,Upa98,Jed99}.
The presence of a large exchange field, $\varepsilon_{\rm{ex}} \gg \Delta$,
suppresses electron-hole correlations in ferromagnets so that the
role of the proximity effect is reduced.
However a separate mechanism has been predicted\cite{deJ95,Fal99a,Fal99b}
to produce a resistance increase in a circuit consisting
of a mono-domain F wire connected to an S electrode instead of an N
electrode.
This is a robust, classical effect which originates from the need to
match a spin-polarized current in the ferromagnet to a spinless
current in the superconductor.

The present work analyses the weak localization correction to the
contact F/S resistance.
In a semiclassical language, weak localization arises from an enhanced
backscattering caused by the quantum interference of pairs of
coherent quasi-particles\cite{Alt82,Ber84,L+R85,C+S86}.
Constructive interference of a quasi-particle paired with its time reversed
partner is destroyed by a magnetic field so we assume that the
width of the F wire, $L_{\perp}$, is small, $L_{\perp} \alt 100$nm,
in order to limit the influence of the intrinsic magnetic field.
Weak localization is affected by the boundary conditions of the system.
Particles that escape into an N electrode suffer dephasing which reduces
the return probability.
However, multiple Andreev reflections from an S electrode may enable a
particle to return coherently.
For a polydomain F wire, with no net average polarization, the return
probability of an F/S system is greater than that of an F/N system\cite{Fal99a}.

In general, the spin channels have different conductivities,
$\sigma_{\alpha} = e^2 \nu_{\alpha} D_{\alpha}$, 
where $\nu_{\alpha}$ and $D_{\alpha}$ are the density of states
and the classical diffusion
coefficient for electrons in the spin state $\alpha$,
$\alpha = (\uparrow, \downarrow )$.
The degree of spin polarization, $\zeta$, is defined as
$\zeta = \left( \sigma_{\uparrow} - \sigma_{\downarrow} \right)
/\left( \sigma_{\uparrow} + \sigma_{\downarrow} \right)$
and we calculate the return probability in an F/S system with an
arbitrary degree of polarization, $0 \leq \zeta \leq 1$.
In addition we introduce a finite spin relaxation length, $L_s$,
shorter than the length, $L$, of the F wire, $L_s \ll L$.
As the spin polarization increases, the majority spin channel experiences
a further increase in return probability (as compared to the F/N case),
whereas the minority spin channel suffers a reduction.
Majority carriers at the superconducting interface cannot find minority
carrier states to Andreev reflect into and are more likely to be
reflected normally whereas minority carriers at the interface
have an enhanced probability to escape.
For $\zeta \rightarrow 1$ the reservoir appears
to be totally insulating to majority spins but normal as far as minority
spins are concerned.

After calculating the correction to the classical diffusion coefficient
that is due to the enhanced return probability,
we find the corresponding change in the spin polarized particle distribution
and we determine the
weak localization correction to the classical resistance.
Generally this is a sum of a bulk term that is independent
of the state of the reservoir and a contact term, $\delta r_c^{FS(FN)}$,
that depends on whether the reservoir is in the superconducting or
the normal state.
We find that $\delta r_c^{FS} > \delta r_c^{FN}$,
so that a change in the state of the reservoir from N to S results in
an increase of the weak localization contribution to the resistance,
\begin{equation}
\delta r_c^{FS} - \delta r_c^{FN} \approx
\gamma \left( \frac{8e^2}{\pi} \right)
\frac{\left( \sigma_{\uparrow} + \sigma_{\downarrow} \right)^2}
{\sigma_{\uparrow}\sigma_{\downarrow} }
\left( R_{\Box} \frac{L_s}{L_{\perp}} \right)^2 ,
\label{dr}
\end{equation}  
where $R_{\Box} = 1/(\sigma_{\uparrow} + \sigma_{\downarrow})L_{\perp}^{d-2}$
is the resistance of a cube of length equal to the width of the
wire $L_{\perp}$.
The combination $R_{\Box}L_s/L_{\perp}$ corresponds to the resistance
of a piece of ferromagnet of length $L_s$.
We evaluate the prefactor, $\gamma$, numerically and find that it is almost
constant, $\gamma \approx 1/2$,
for all experimentally relevant values of spin polarization.

%
%
\section{Classical Resistance}
%
We begin by briefly describing the approach of
Refs.~\onlinecite{Fal99a,Fal99b} which enables one to calculate
the classical contact resistance of an F/S interface and we refer
the reader to those papers for further details.
We consider a single domain, disordered ferromagnetic wire of length
$L$ as depicted in Fig.~1.
The coordinate $x$ is used to describe the position along the
wire and throughout the paper we consider there to be a
normal reservoir at the left hand side, $x= -L$, whereas
the reservoir at the right hand side, $x= 0$, is either N or S.
The resistance of the disordered F wire is determined by solving
diffusion equations for the isotropic part of the electron
distribution function
$n_{\alpha} (\varepsilon,x) = \int d\Omega_{\bf p}
n_{\alpha} ({\bf p},x) .$
Using electron-hole symmetry, it is possible to consider only the
symmetrised function
$N_{\alpha} (\varepsilon,x) = [n_{\alpha} (\varepsilon,x)
+ n_{\alpha} (-\varepsilon,x)]/2 ,$
where $\varepsilon$ is the energy determined with respect to the chemical
potential in the S (N) reservoir.
The current density in any given spin channel is related to the
spatial derivative of the distribution function as
\begin{equation}
j_{\alpha} = \sigma_{\alpha} \int_{0}^{\infty} d\varepsilon \,
\partial_x N_{\alpha} (\varepsilon,x) .
\label{classicalj}
\end{equation}
The distribution function obeys a differential equation
describing diffusion in the ferromagnetic wire,
\begin{equation}
D_{\alpha} \partial_x^2 N_{\alpha} (\varepsilon,x)
= w_{\uparrow\downarrow} \nu_{\overline{\alpha}}
\left[
N_{\alpha} (\varepsilon,x) - N_{\overline{\alpha}} (\varepsilon,x)
\right] ,
\label{de1}
\end{equation}
where $\overline{\alpha} = (\downarrow ,\uparrow )$
for $\alpha = (\uparrow, \downarrow )$.
This equation may be expressed conveniently as
\begin{eqnarray}
\partial_x^2 \left[
\sigma_{\uparrow} N_{\uparrow} + \sigma_{\downarrow} N_{\downarrow}
\right]  &=& 0 ,\label{de2} \\
\left( \partial_x^2 - L_{s}^{-2} \right)
\left[
N_{\uparrow} - N_{\downarrow}
\right]  &=& 0 .\label{de3}
\end{eqnarray}
Spin relaxation, which may arise from spin-orbit scattering or
spin-flip scattering at defects, is described by the right-hand
side of Eq.~(\ref{de1}).
One may define an effective spin relaxation rate, $L_s$,
as
$L_s^{-2} = w_{\uparrow\downarrow}
[\nu_{\uparrow} / D_{\downarrow} + \nu_{\downarrow} / D_{\uparrow}]$
which accounts for the relaxation of the difference between the spin channels.
In a similar way, the spin relaxation length in an individual channel,
$L_{\alpha}$, is defined as
$
L_{\alpha}^{-2} = w_{\uparrow\downarrow}
\nu_{\overline{\alpha}} / D_{\alpha}
$
such that $L_s^{-2} = L_{\uparrow}^{-2} + L_{\downarrow}^{-2}$.
An equivalent expression is
$
L_{\alpha}^2 = L_s^2 
(\sigma_{\alpha} + \sigma_{\overline{\alpha}})
/{\sigma_{\overline{\alpha}}}
$

The above equations should be complemented by
four boundary conditions, two at each end of the ferromagnetic wire.
The boundary conditions at the left hand side of the wire are given
by the equilibrium distribution of electrons in the normal
reservoir,
\begin{equation}
N_{\alpha} (\varepsilon, -L) =
\frac{1}{2} \left[
n_{T} (\varepsilon -eV) + n_{T} (-\varepsilon -eV)
\right] ,
\label{bc1}
\end{equation}
such that the distribution of up and down spins is equal.
The boundary conditions at the right hand side of the ferromagnetic
wire at the superconducting reservoir have been discussed in detail in
Refs.~\onlinecite{Fal99a,Fal99b}.
For quasiparticles with energies below the superconducting gap they may
be written as
\begin{eqnarray}
\left. \sigma_{\uparrow} \partial_x N_{\uparrow} \right|_0
&=& 
\left. \sigma_{\downarrow} \partial_x N_{\downarrow} \right|_0,
\label{bc2} \\
N_{\uparrow} (\varepsilon, 0) + N_{\downarrow} (\varepsilon, 0)
&=&
\left[
n_{T} (\varepsilon ) + n_{T} (-\varepsilon )
\right] .
\label{bc3}
\end{eqnarray}
The first of these, Eq.~(\ref{bc2}), describes Andreev reflection such that
the spin-up current must equal the spin-down current and the
second, Eq.~(\ref{bc3}), states that the sum of the distributions
is equal to the equilibrium value.
On solving the above differential equations with the appropriate
boundary conditions it is possible to calculate the current density
in the wire.
In particular, the spatial derivative of the distribution is
\begin{eqnarray}
\partial_x N_{\alpha} (\varepsilon, x)
=&&
\frac{(1-2 N_{L})}{2\Gamma L} \nonumber \\
&& \!\!\!\!\!\! \!\!\!\!\!\! \times \left[
1 +
\frac{(\sigma_{\overline{\alpha}} - \sigma_{\alpha} )}
{2 \sigma_{\alpha}}
\frac{\cosh [(x+L)/L_s ]}{\cosh (L/L_s)}
\right] ,
\label{dN0}
\end{eqnarray}
where
\begin{equation}
\Gamma = 1 + \frac{(\sigma_{\uparrow} - \sigma_{\downarrow} )^2}
{4\sigma_{\uparrow}\sigma_{\downarrow}}
\frac{L_s}{L}
\tanh (L/L_s) ,
\label{gam}
\end{equation}
and $N_L \equiv N_{\alpha} (\varepsilon, -L)$.

The effect of the superconducting reservoir, for strong spin relaxation
$L \gg L_s$, is expressed in terms of a contact resistance $r_c^{FS}$ so that
the total classical resistance, $R_{cl}$, may be written as a sum
of resistances in series
\begin{equation}
R_{cl} = \frac{L}{L_{\perp}} R_{\Box} + r_c^{FS} ,
\end{equation}
\begin{equation}
r_c^{FS}
\approx R_{\Box} \frac{L_{s}}{L_{\perp}}
\frac{\left( \sigma_{\uparrow} - \sigma_{\downarrow} \right)^2}
{4 \sigma_{\uparrow}\sigma_{\downarrow} } .
\label{rc}
\end{equation}
For a normal reservoir on the right hand side, the boundary conditions
are similar to those in Eq.~(\ref{bc1}), namely
$N_{\alpha} (\varepsilon, 0) =
\left[
n_{T} (\varepsilon ) + n_{T} (-\varepsilon )
\right] /2$.
In this case, there is no contact resistance.
A fall in temperature, precipitating a change in state of the
reservoir from N to S, would therefore result in a resistance
increase of the circuit.
This is a robust classical effect which originates from the need to
match a spin-polarized current in the ferromagnet to a spinless
current in the superconductor.
It is specific to mono-domain wires since in poly-domain wires
the tranport properties of spin-up and spin-down particles do not
differ and, on the average, the current is not spin-polarized.

%
\section{Weak localization correction to the contact resistance}
\label{Sec:wl}
%
The weak localization correction to the classical diffusion coefficient,
$\delta D_{\alpha}$,
is related to a two particle Green's function known
as the Cooperon propagator\cite{Alt82,Ber84,L+R85,C+S86}.
The Cooperon consists of a quasi-particle following a diffusive path
that interferes with a quasi-particle
traversing the same path in the opposite direction.
When the particles follow a closed path, the interference results in an
enhanced return probability.
In a ferromagnet, singlet and $S_z = 0$ triplet Cooperons are suppressed
since the Fermi momentum for spin-up and spin-down
particles is different, producing a difference in the phase accumulated
by the particles along the path.
However, there is no such suppression of the phase correlation in
the triplet channel where same spin particles are paired.
Thus we assume that
the return probability is given by
the triplet Cooperon propagator,
$C_{\alpha \alpha} (x ,x^{\prime})$,
for equal spatial coordinates,\cite{Alt82,Fal99a}
\begin{equation}
\frac{\delta D_{\alpha} (x)}{D_{\alpha}^{0}}
\approx
- \frac{8}{\pi L_{\perp}^2} \,
C_{\alpha \alpha} (x ,x)
,\label{retprob}
\end{equation}
which obeys a diffusion equation in the disordered ferromagnet,
\begin{equation}
\left( -\nu_{\alpha} D_{\alpha} \partial_x^2 +
\nu_{\alpha} \tau_{\alpha}^{-1} \right)
C_{\alpha \alpha} (x ,x^{\prime})
=  \delta \left( x - x^{\prime} \right)  .
\label{coopde}
\end{equation}
Here the spin relaxation time, $\tau_{\alpha}$, is related to the spin
relaxation length in the spin channel $\alpha$,
$\tau_{\alpha} = L_{\alpha}^2 /D_{\alpha}$.

We stress that the existence of a Cooperon is not a result
of the proximity effect as we do not consider triplet pairing
induced in the ferromagnet by the presence of the superconductor.
Nevertheless,
weak localization is affected by the boundary conditions of the system.
A schematic of the change in the boundary conditions of the Cooperon
at the F/S interface as compared to the F/N interface is shown
in Fig.~2.
A process which pairs two spin-up quasi-particles is depicted
in Fig.~2a.
When the reservoir is in the N state, particles escaping into it suffer
dephasing, thus destroying the Cooperon.
At the left hand reservoir ($x=-L$), which is in the N state,
we therefore have
\begin{equation}
C_{\uparrow\uparrow} (-L,x^{\prime}) =
C_{\downarrow\downarrow} (-L,x^{\prime}) = 0
. \label{coopbc1}
\end{equation}
When the right hand reservoir ($x=0$) is in the S state we apply
\begin{eqnarray}
C_{\uparrow\uparrow} (0,x^{\prime}) &=&
{\rm e}^{i\chi} C_{\downarrow\downarrow} (0,x^{\prime})
, \label{coopbc2} \\
\sigma_{\uparrow} \left.
\partial_x C_{\uparrow\uparrow} (x,x^{\prime}) \right|_{x=0}
&=&
- \sigma_{\downarrow} {\rm e}^{i\chi}\left.
\partial_x C_{\downarrow\downarrow} (x,x^{\prime}) \right|_{x=0}
, \label{coopbc3}
\end{eqnarray}
where $\chi$ is a phase accumulated on reflection from the superconductor.
These boundary conditions, similar to those for the density
Eqs.(\ref{bc2},\ref{bc3}), account for Andreev  reflection.
Multiple Andreev  reflections at an S electrode may enable a
particle to return coherently to the original place in the F wire
with the same spin polarization.
Fig.~2b shows a process whereby a pair of spin-up quasi-particles
are Andreev reflected as spin-down quasi-holes which are
subsequently Andreev reflected as spin-up quasi-particles.
The calculation of $\delta D_{\alpha}$ by solving the Cooperon
diffusion equation with the above boundary
conditions is described in Appendix~\ref{app:A}.

The weak localization correction to the diffusion coefficient,
$\delta D_{\alpha}$,
leads to a correction to the current density,
$\delta j_{\alpha} = e^2 \nu_{\alpha} \delta D_{\alpha}
\int_{0}^{\infty} d\varepsilon \,
\partial_x N_{\alpha} (\varepsilon,x)$.
However, $\delta D_{\alpha}$
also produces a correction to the particle distribution function.
We take this into account perturbatively by assuming that the
correction, $\delta D_{\alpha}$, to the classical diffusion
coefficient, $D_{\alpha}^{0}$, is small and that there is a corresponding
small correction, $\delta N_{\alpha}$, to the classical particle
distribution, $N_{\alpha}^{0}$.
Expressions for the total diffusion coefficient,
$D_{\alpha} = D_{\alpha}^{0} + \delta D_{\alpha}$,
and the total particle distribution,
$N_{\alpha} = N_{\alpha}^{0} + \delta N_{\alpha}$,
are substituted into the previous differential equations
describing diffusion in the ferromagnet,
Eqs.~(\ref{de2},\ref{de3}),
and the boundary conditions,
Eqs.~(\ref{bc1},\ref{bc2},\ref{bc3}),
to provide new equations for $\delta N_{\alpha}$.
Hence the differential equations for the correction to the distribution
function,
$\delta N_{\alpha}$, in terms of the correction to the classical
diffusion coefficient, $\delta D_{\alpha}$, are
\begin{eqnarray}
\partial_x^2 \left[
\sigma_{\uparrow} \delta N_{\uparrow} + \sigma_{\downarrow} \delta N_{\downarrow}
\right]  &=&
- \partial_x \left[
\delta {\cal J}_{\uparrow} + \delta {\cal J}_{\downarrow} 
\right]
,      \label{wlde1} \\
\left( \partial_x^2 - L_{s}^{-2} \right)
\left[
\delta N_{\uparrow} - \delta N_{\downarrow}
\right]  &=&
- \partial_x \left[
\frac{\delta {\cal J}_{\uparrow}}{\sigma_{\uparrow}}
- \frac{\delta {\cal J}_{\downarrow}}{\sigma_{\downarrow}}
\right]
, \label{wlde2}
\end{eqnarray}
where
$\delta {\cal J}_{\alpha} (x) = e^2 \nu_{\alpha} \delta D_{\alpha} (x)
\, \partial_x N_{\alpha}^{0} (x)$.
The boundary conditions at the ferromagnetic reservoir on the left hand
side are
\begin{equation}
\delta N_{\uparrow} (\varepsilon, -L)
= \delta N_{\downarrow} (\varepsilon, -L)
= 0
. \label{wlbc1}
\end{equation}
whereas the boundary conditions at the superconducting reservoir on
the right hand side are
\begin{eqnarray}
\left. \sigma_{\uparrow} \partial_x \delta N_{\uparrow} \right|_0
- \left. \sigma_{\downarrow} \partial_x \delta N_{\downarrow} \right|_0
&=&
\delta {\cal J}_{\downarrow}
- \delta {\cal J}_{\uparrow} ,
\label{wlbc2} \\
\delta N_{\uparrow} (\varepsilon, 0)
+ \delta N_{\downarrow} (\varepsilon, 0)
&=& 0 .
\label{wlbc3}
\end{eqnarray}

General solutions to the differential equations,
Eqs.~(\ref{wlde1},\ref{wlde2}),
are
\begin{equation}
\sigma_{\uparrow} \delta N_{\uparrow} + \sigma_{\downarrow} \delta N_{\downarrow}
=
Ux + V
- \int^{y=x} \!\!\!
\left[
\delta {\cal J}_{\uparrow} (y) + \delta {\cal J}_{\downarrow} (y)
\right] dy 
,      \label{gs1}
\end{equation}
and
\begin{eqnarray}
\delta N_{\uparrow} - \delta N_{\downarrow}
&=&
We^{-x/L_s} + Ye^{+x/L_s}   \nonumber \\
&& \!\!\!\!\!\!\!\!\!\!\!\!\!\!\!\!\!\!\!\!\!\!\!\!
- \int^{y=x} \!\! \left[
\frac{\delta {\cal J}_{\uparrow} (y)}{\sigma_{\uparrow}}
- \frac{\delta {\cal J}_{\downarrow} (y)}{\sigma_{\downarrow}}
\right]
\cosh \left( \frac{x-y}{L_s} \right)  dy 
 .  \nonumber \\
\label{gs2}
\end{eqnarray}
where $U$,$V$,$W$, and $Y$ are unknown factors to be determined by the four
boundary conditions.
It is thus possible to evaluate $\delta N_{\alpha}$ in terms of the
correction to the diffusion coefficient, $\delta D_{\alpha}$,
and the classical part of the distribution, $N_{\alpha}^{0}$.

Taking into account both $\delta D_{\alpha}$ and $\delta N_{\alpha}$,
the weak localization correction to the current density summed over both
spin channels is
\begin{equation}
\delta j = \int_{0}^{\infty} d\varepsilon \,
\left[
\delta {\cal J}_{\uparrow}
+ \delta {\cal J}_{\downarrow}
+ \sigma_{\uparrow} \partial_x \delta N_{\uparrow}
+ \sigma_{\downarrow} \partial_x \delta N_{\downarrow}
\right]
. \label{wlj}
\end{equation}
This is determined by the factor $U$ in the general solution
Eq.~(\ref{gs1}).
Using the boundary conditions, Eqs.~(\ref{wlbc1},\ref{wlbc2},\ref{wlbc3}),
to evaluate $U$, the weak localization correction to the current density
may be expressed as
\begin{eqnarray}
\delta j &=&
\int_{0}^{\infty} \frac{d\varepsilon}{\Gamma}
\int_{-L}^{0} \frac{dx}{L}
\Bigg\{
\left[
\delta {\cal J}_{\uparrow}
+ \delta {\cal J}_{\downarrow}
\right]  + 
 \nonumber  \\
&& 
- \frac{(\sigma_{\uparrow}-\sigma_{\downarrow} )}{2}
\left[
\frac{\delta {\cal J}_{\uparrow}}{\sigma_{\uparrow}}
- \frac{\delta {\cal J}_{\downarrow}}{\sigma_{\downarrow}}
\right]
\frac{\cosh \left[ (x+L)/L_s \right]}
{\cosh \left[ L/L_s \right]}
\Bigg\}
.     \nonumber \\
\end{eqnarray}
The term in brackets contains spatial dependences arising from
the $\cosh$ term and from
$\delta {\cal J}_{\alpha} (x) = e^2 \nu_{\alpha} \delta D_{\alpha} (x)
\, \partial_x N_{\alpha}^{0} (x)$,
where $\partial_x N_{\alpha}^{0} (x)$ depends on
$\cosh [(x+L)/L_s ]$,
Eq.(\ref{dN0}), and $\delta D_{\alpha} (x)$ is
given in Appendix~\ref{app:A}.
Taking these terms into account, it is straightforward to perform
the integration with respect to the coordinate $x$.
The final step is to evaluate $\delta D_{\alpha} (x)$ numerically
as detailed in Appendix~\ref{app:A}.
For a long wire, we express the weak localization correction 
as a sum of resistances,
\begin{equation}
\delta R = \delta R^{\rm{B}} +  \delta r_c^{FS}
\label{dR}
\end{equation}  
where $\delta R^{\rm{B}}$ is the bulk contribution,
$\delta R^{\rm{B}} = (8e^2/\pi )R_{\Box}^2 L(L_{\uparrow} + L_{\downarrow})
/L_{\perp}^2$ and $\delta r_c^{FS}$ is a contact term,
\begin{equation}
\delta r_c^{FS} \approx -(1-\gamma )\left( \frac{8e^2}{\pi} \right)
\frac{\left( \sigma_{\uparrow} + \sigma_{\downarrow} \right)^2}
{\sigma_{\uparrow}\sigma_{\downarrow} }
\left( R_{\Box} \frac{L_s}{L_{\perp}} \right)^2 ,
\label{dr2}
\end{equation}  
The parameter $\gamma$ is plotted as a function of the degree of polarization,
$\zeta$, in Fig.~3.
The solid line is $\gamma$ for $L/L_s = 20$
whereas the dashed line is the contribution of the first two terms
in Eq.~(\ref{wlj}) only, neglecting the influence of density
variations.
It is worth noting that the analytic result for an F/N system is
$\gamma = 0$ and for a polydomain wire connected
to an S reservoir it is $\gamma = 1/2$\cite{Fal99a}.
In these cases, $\delta N = 0$.
For moderate values of $\zeta$, we find that there is no large change in
$\gamma$, $\gamma \approx 1/2$, and the influence of
density variations is small.
However, for very large spin polarization, $\zeta \agt 0.8$, $\gamma$
increases dramatically and the role of density variations is
vital.
Drawing on the interpretation of the results for the return probability
mentioned in Appendix~\ref{app:A},
we naively estimate the resistance of a system with $\zeta = 1$ by
considering the majority (up) spin channel to behave as if connected to
an insulating barrier and the minority channel to behave as if connected
to a normal barrier.
Such a procedure gives
$\gamma \sim 1 - \sigma_{\downarrow}/(\sigma_{\uparrow} + \sigma_{\downarrow})
\sim 1$.
This rough estimation of the value of $\gamma$ at $\zeta = 1$ appears
to be in agreement with Fig.~3.
However, some notes of caution about the results for large polarization
need to made.
Firstly, the graphs are shown as a function of $\zeta$ for fixed $L/L_s = 20$,
whereas the spin relaxation lengths, $L_{\alpha}$, in
the individual spin channels depend on $\zeta$.
In particular, $L_{\uparrow} \gg L_s$ for large $\zeta$.
Secondly, the error in the numerical evaluation of the diffusion coefficient
increases rapidly as $\zeta \rightarrow 1$.
Finally, values of $\zeta \agt 0.8$ are unlikely to be realised in
real materials and, even if they were, the intrinsic magnetic field
in such materials would destroy any quantum interference.

%
%
\section{Conclusion}
%
The weak localization correction to the classical diffusion
coefficient, $\delta D$, is dependent on polarization,
with majority spins more likely
to be reflected from the F/S interface than minority spins.
Taking into account the change in the spin polarized particle distribution
in the F wire arising from $\delta D$, we found that the weak localization
correction to the contact resistance is related to the square of the resistance
of a piece of F wire with length equal to the spin relaxation length
(see Eq.~(\ref{dr})) with a numerical prefactor that is almost constant
for all experimentally relevant values of spin polarization.

\acknowledgments
The authors are grateful to D.~E.~Khmelnitskii, B.~Pannetier, V.~T.~Petrashov,
and I.~A.~Sosnin for useful discussions.
This work was supported by EPSRC. 
%
%
%
%
\appendix
%
%
%
\section{Correction to the diffusion coefficient}
\label{app:A}
%
This appendix describes the evaluation of the 
weak localization correction to the classical diffusion coefficient,
$\delta D_{\alpha}$, Eq.~(\ref{retprob}).
In order to calculate the Cooperon, Eq.~(\ref{coopde}),
we consider spinor eigenvectors with components $\psi_{n\alpha} (x)$
that obey the following diffusion equations:
\begin{eqnarray}
\left( -\nu_{\uparrow} D_{\uparrow} \partial_x^2
+ \nu_{\uparrow}\tau_{\uparrow}^{-1} \right)
\psi_{n\uparrow} (x)
&=& \lambda_{n} \, \psi_{n\uparrow} (x)  , \label{Acoopde1} \\
\left( -\nu_{\downarrow} D_{\downarrow} \partial_x^2
+ \nu_{\downarrow}\tau_{\downarrow}^{-1} \right)
\psi_{n\downarrow} (x)
&=& \lambda_{n} \, \psi_{n\downarrow} (x) 
, \label{Acoopde2}
\end{eqnarray}
such that the Cooperon $C_{\alpha\alpha}$ is given by
\begin{equation}
C_{\alpha\alpha} (x,x) = \sum_{n \neq 0}^{\infty} 
\frac{\left| \psi_{n \alpha} (x) \right|^2}{\lambda_{n}} .
\label{coopdef}
\end{equation}
The momenta in the spin channels, $Q_{n\alpha}$, are related,
$\nu_{\uparrow} D_{\uparrow} Q_{n\uparrow}^2 +
\nu_{\uparrow}\tau_{\uparrow}^{-1} =
\nu_{\downarrow} D_{\downarrow} Q_{n\downarrow}^2 +
\nu_{\downarrow}\tau_{\downarrow}^{-1} \equiv \lambda_{n} .$
At the left hand reservoir ($x=-L$) the Cooperon is zero since a particle
that escapes into it suffers dephasing, Eq.~(\ref{coopbc1}),
\begin{equation}
\psi_{n\uparrow} (-L) = \psi_{n\downarrow} (-L) = 0
. \label{Acoopbc1}
\end{equation}
whereas when the reservoir on the right hand ($x=0$) is a
superconductor, the boundary conditions
Eqs.~(\ref{coopbc2},\ref{coopbc3}) give
\begin{eqnarray}
\psi_{n\uparrow} (0) &=& {\rm e}^{i\chi} \psi_{n\downarrow} (0)
, \label{Acoopbc2} \\
\sigma_{\uparrow} \left. \partial_x \psi_{n\uparrow} (x) \right|_{x=0}
&=&
- \sigma_{\downarrow} {\rm e}^{i\chi}\left.
\partial_x \psi_{n\downarrow} (x) \right|_{x=0}
. \label{Acoopbc3}
\end{eqnarray}
The spinor eigenvectors also obey a normalisation condition,
$\int_{-L}^{0}(|\psi_{n\uparrow} (x)|^2 + |\psi_{n\downarrow} (x)|^2)dx = 1$.
We find the general solutions to the diffusion equations and
use the boundary conditions at the left hand lead ($x = -L$),
Eq.~(\ref{Acoopbc1}), to give
\begin{equation}
\psi_{n\alpha} (x) = {\cal N}_{n\alpha} \sin \left[ Q_{n\alpha} (x+L) \right]
\label{efn}
\end{equation}
On substituting into the boundary conditions on the right hand side ($x=0$), 
we find
\begin{eqnarray}
{\cal N}_{n\uparrow} \sin \left[ Q_{n\uparrow} L \right]
&=&
{\rm e}^{i\chi} {\cal N}_{n\downarrow} \sin \left[ Q_{n\downarrow} L \right]
,  \label{coopbc2a} \\
\sigma_{\uparrow}  Q_{n\uparrow} {\cal N}_{n\uparrow}
\cos \left[ Q_{n\uparrow} L \right]
&=&
- {\rm e}^{i\chi} \sigma_{\downarrow} Q_{n\downarrow} {\cal N}_{n\downarrow} 
\cos \left[ Q_{n\downarrow} L \right]
. \label{coopbc1a}
\end{eqnarray}
Eliminating the normalisation constants leaves an equation
for determining the eigenvalues,
\begin{eqnarray}
\sigma_{\downarrow} Q_{n\downarrow}
&&\cos \left[ Q_{n\downarrow} L \right] 
\sin \left[ Q_{n\uparrow} L \right]
+ \nonumber \\
&&\sigma_{\uparrow} Q_{n\uparrow} \cos \left[ Q_{n\uparrow} L \right]
\sin \left[ Q_{n\downarrow} L \right]
=0 \, . \label{fq}
\end{eqnarray}  
We solve this equation numerically to find the eigenvalues and
determine their contribution to the Cooperon, Eq.~(\ref{coopdef}).

As an example we present results for the calculation of the spatially
averaged return probability,
$\overline{\delta D}_{\alpha} = \int_{-L}^{0} 
\delta D_{\alpha} (x)
(dx/L),$
for arbitrary spin polarization in the ferromagnet.
For long wires, $L \gg L_S$, we express it as
\begin{equation}
\overline{\delta D}_{\alpha}
\approx
{\delta D}_{\alpha}^{\rm{B}}
\left( 1 - \eta_{\alpha}\frac{L_{\alpha}}{L} \right)
, \label{dD}
\end{equation}  
where ${\delta D}_{\alpha}^{\rm{B}}$ is the bulk contribution,
${\delta D}_{\alpha}^{\rm{B}} =
- 8L_{\alpha} / \pi \nu_{\alpha} L_{\perp}^2$.
The numerical coefficient $\eta_{\alpha}$ depends on the state of the reservoir
and the degree of polarization in the ferromagnet.
We compare with analytic results obtained in certain limits
\cite{Fal99a}.
When the reservoir on the right hand side of the wire is in the
normal state, the boundary conditions are the same as those in
the left hand reservoir and $\eta_{\alpha} = 1$.
For a poly-domain ferromagnetic wire connected to an S reservoir,
where the classical contact resistance Eq.~(\ref{rc}) gives
no effect, $\eta_{\alpha} = 1/2$.
The second term in the above equation represents Cooperon decay
due to the probability of particle escape into the right hand
reservoir.
Multiple Andreev reflection at the boundary, illustrated in a sketch in
Fig.~2b, causes a factor of two
reduction in the case of a polydomain ferromagnetic
wire connected to a superconducting reservoir.
In the case of a ferromagnetic wire connected to an insulating
reservoir, $\eta_{\alpha} = 0$, since the probability of particle
escape into the right hand reservoir is totally supressed.

In Fig.~4 we plot
$\eta_{\alpha}$ for a ferromagnetic wire with arbitrary
polarization connected to a superconducting reservoir.
It is shown for each spin channel separately and we assume that
the up-spins are the majority spins, $\sigma_{\uparrow} \geq \sigma_{\downarrow}$.
For zero spin polarization,
$\eta_{\uparrow} = \eta_{\downarrow} = 1/2$,
as described above.
On increasing the spin polarization, $\eta_{\uparrow} < 1/2$
whereas $\eta_{\downarrow} > 1/2$.
Majority carriers at the superconducting interface cannot find minority
carrier states to Andreev reflect into and are normally reflected which increases
the probability to return whereas minority carriers at the interface
have an enhanced probability to escape.
In the limit of total spin polarization, $\eta_{\uparrow} \rightarrow 0$
and $\eta_{\downarrow} \rightarrow 1$ which means that the reservoir appears
to be totally insulating to majority spins but normal as far as minority
spins are concerned.

%

\end{multicols}
\newpage

\begin{figure}
\vspace{0.3cm}
\hspace{0.02\hsize}
\epsfxsize=0.9\hsize
\epsffile{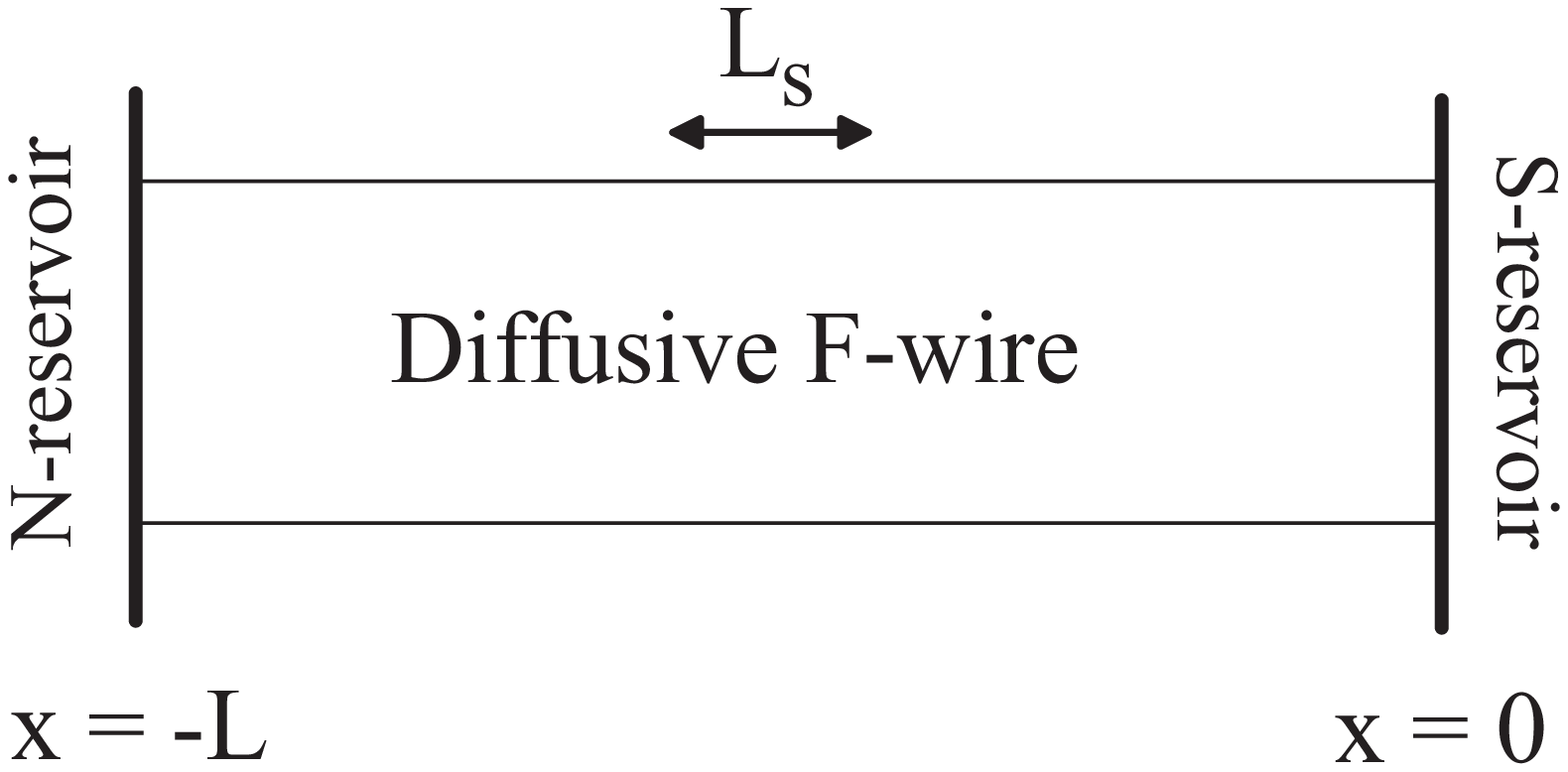}
\vspace{0.3cm}
 
\refstepcounter{figure}
\label{fig:1}
{\setlength{\baselineskip}{10pt} FIG.\ 1.
Schematic of the ferromagnetic wire connected to a
normal (N) reservoir on the left hand side, $x=-L$, and
a superconducting (S) reservoir on the right hand side, $x=0$.
}
\end{figure}

\begin{figure}
\vspace{0.3cm}
\hspace{0.02\hsize}
\epsfxsize=0.9\hsize
\epsffile{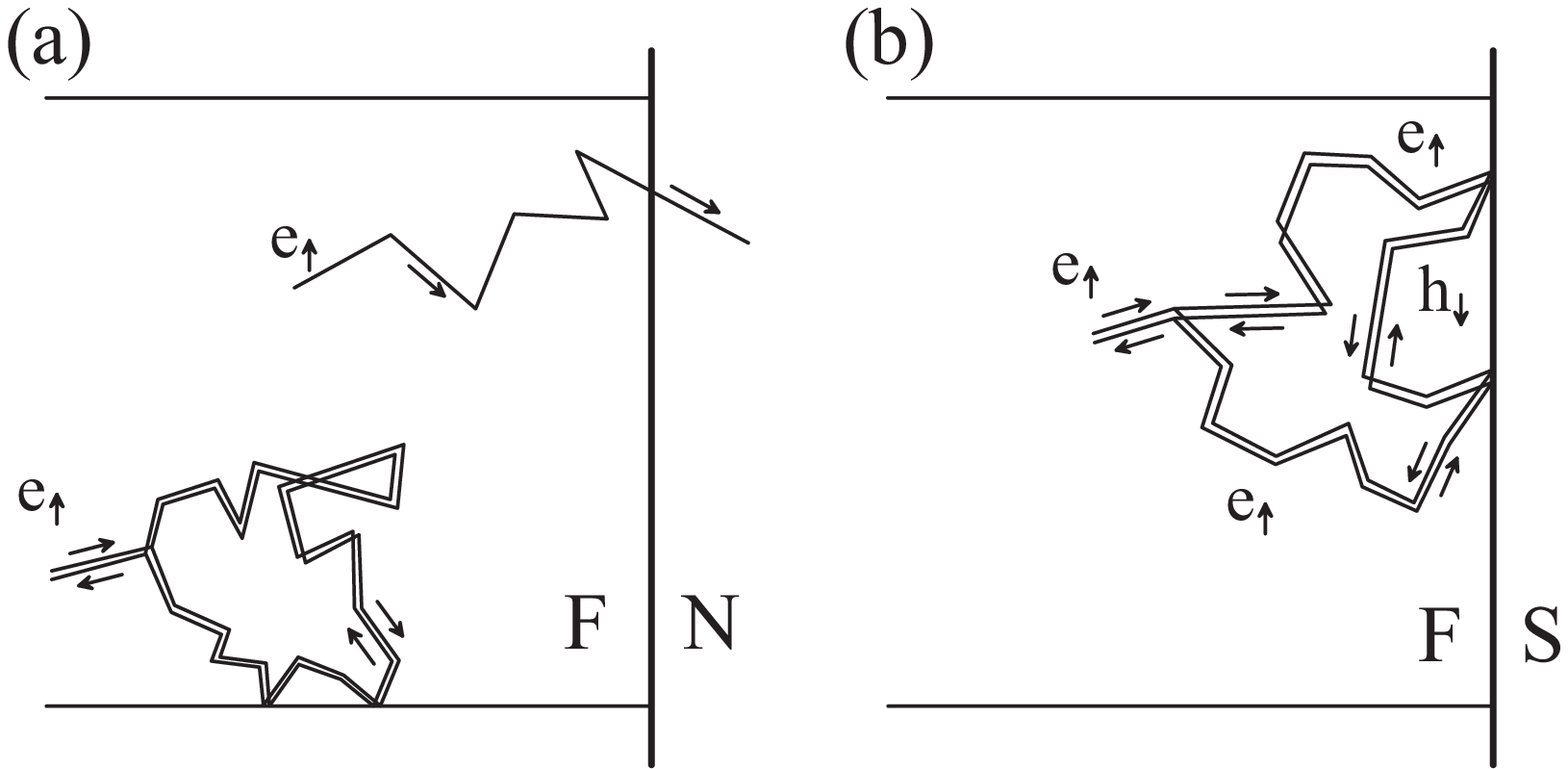}
\vspace{0.3cm}
 
\refstepcounter{figure}
\label{fig:2}
{\setlength{\baselineskip}{10pt} FIG.\ 2.
(a) Cooperon decay at the F/N boundary due to electron escape into
the N reservoir.
(b) Multiple Andreev reflection at the S reservoir changes the
Cooperon boundary conditions at the interface.
}
\end{figure}

\begin{figure}
\vspace{0.3cm}
\hspace{0.02\hsize}
\epsfxsize=0.9\hsize
\epsffile{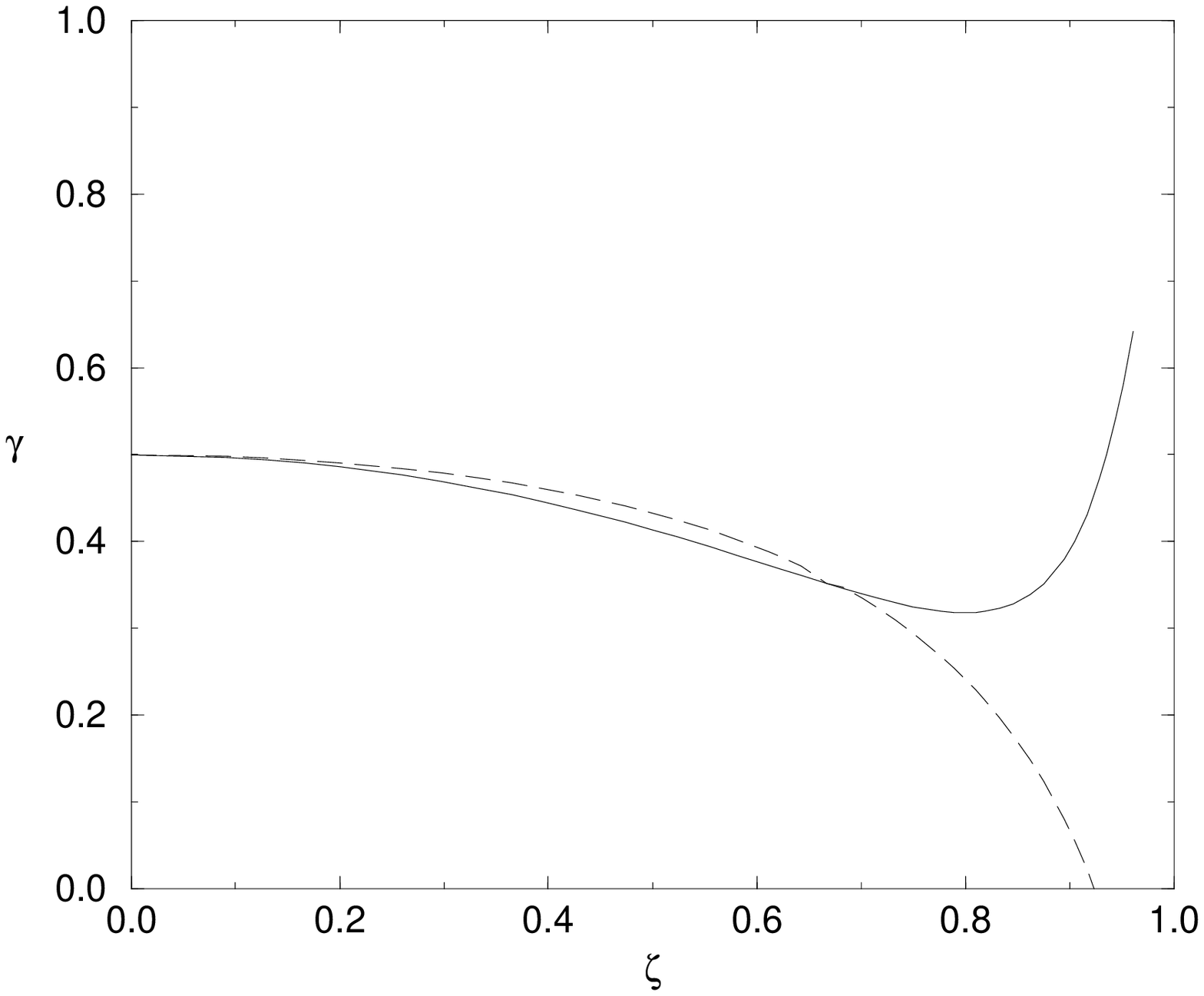}
\vspace{0.3cm}
 
\refstepcounter{figure}
\label{fig:3}
{\setlength{\baselineskip}{10pt} FIG.\ 3.
The prefactor, $\gamma$, of the increase, $\delta r_c^S - \delta r_c^N$,
(Eq.~(\ref{dr})) in the weak localization contribution to the resistance
as the reservoir changes from the normal state to superconducting,
plotted as a function of the degree of polarization, $\zeta$,
for $L/L_s = 20$.
The solid line is the total weak localization contribution
whereas the dashed line does not include the contribution of
density variations.
The error increases as $\zeta \rightarrow 1$ (see comments at the end
of section~\ref{Sec:wl}).
}
\end{figure}

\begin{figure}
\vspace{0.3cm}
\hspace{0.02\hsize}
\epsfxsize=0.9\hsize
\epsffile{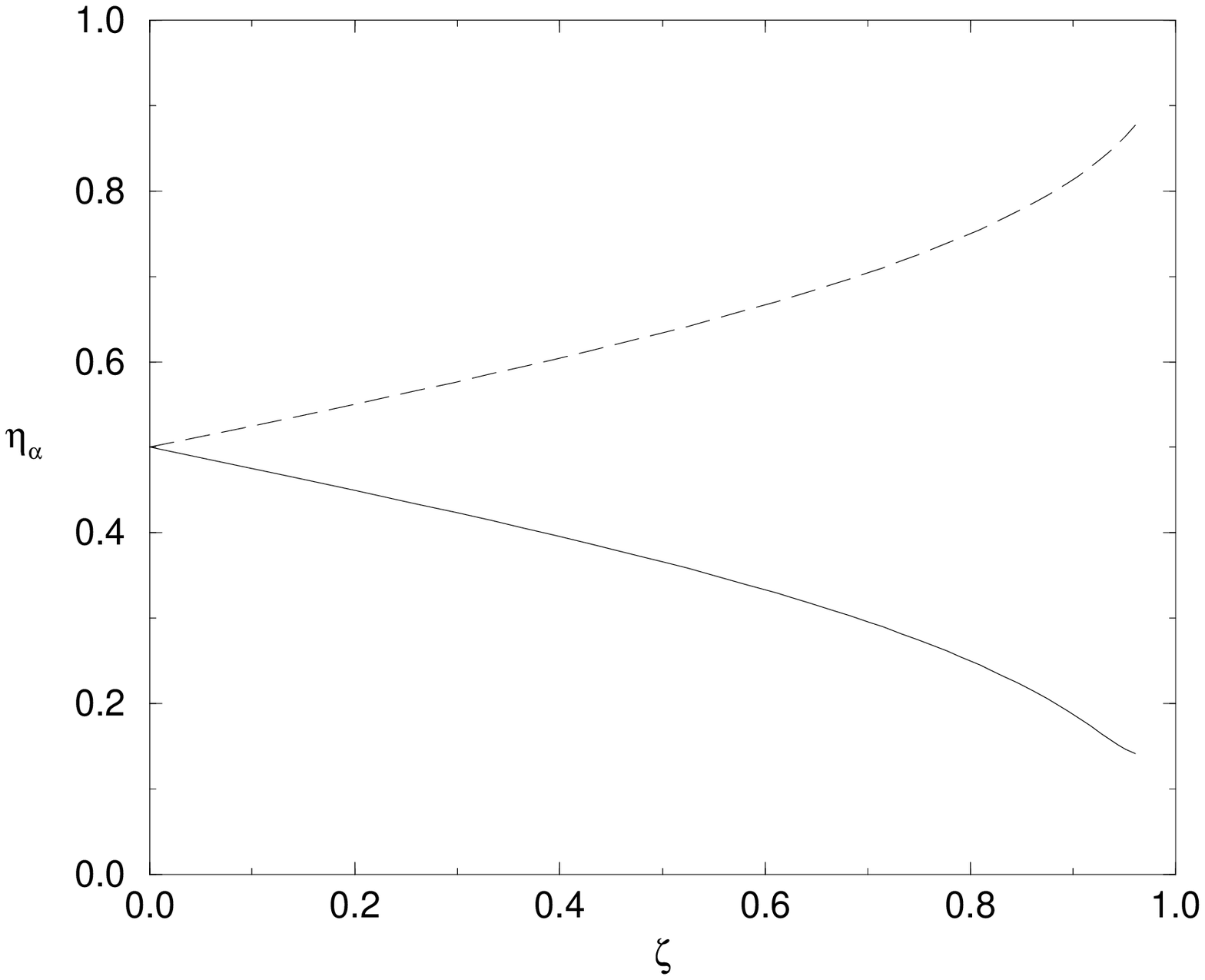}
\vspace{0.3cm}
 
\refstepcounter{figure}
\label{fig:4}
{\setlength{\baselineskip}{10pt} FIG.\ 4.
The prefactor, $\eta_{\alpha}$, of the contact term in
the spatially averaged return probability, $\overline{\delta D}_{\alpha}$,
(Eq.~(\ref{dD})) for the F/S system,
plotted as a function of the polarization, $\zeta$, for $L/L_s = 20$.
The solid line is for the majority spin channel whereas the dashed
line is for minority spins.
}
\end{figure}

\end{document}